\documentclass[12pt]{iopart}
\usepackage{graphicx, multirow}

\begin{document}

\title[Qubit compatible superconducting interconnects]{Qubit compatible superconducting interconnects}

\author{B. Foxen,$^{1}$ J.Y. Mutus,$^{2}$ E. Lucero,$^{2}$ R. Graff,$^{2}$ A. Megrant,$^{2}$ Yu Chen,$^{2}$ C. Quintana,$^{1,2}$ B. Burkett,$^{2}$ J. Kelly,$^{2}$ E. Jeffrey,$^{2}$ Yan Yang,$^{3}$ Anthony Yu,$^{3}$ K. Arya,$^{2}$ R. Barends,$^{2}$ Zijun Chen,$^{1}$ B. Chiaro,$^{1}$ A. Dunsworth,$^{1}$ A. Fowler,$^{2}$ C. Gidney,$^{2}$ M. Giustina,$^{2}$ T. Huang,$^{2}$ P. Klimov,$^{2}$ M. Neeley,$^{2}$ C. Neill,$^{1}$ P. Roushan,$^{2}$ D. Sank,$^{2}$ A. Vainsencher,$^{2}$ J. Wenner,$^{1}$ T.C. White,$^{2}$ 
and John M. Martinis$^{1,2,b)}$}

\address{$^1$ Department of Physics, University of California, Santa Barbara, CA 93106-9530}
\address{$^2$ Google, Santa Barbara, CA 93117}
\address{$^3$ Google, Mountain View, CA 94043}
\ead{$^b$ martinis@physics.ucsb.edu}
\vspace{10pt}
% \begin{indented}
% % \item[]August 2017
% \end{indented}

\begin{abstract}
We present a fabrication process for fully superconducting interconnects compatible with superconducting qubit technology. These interconnects allow for the three dimensional integration of quantum circuits without introducing lossy amorphous dielectrics. They are composed of indium bumps several microns tall separated from an aluminum base layer by titanium nitride which serves as a diffusion barrier. We measure the whole structure to be superconducting (transition temperature of 1.1\,K), limited by the aluminum. These interconnects have an average critical current of 26.8\,mA, and mechanical shear and thermal cycle testing indicate that these devices are mechanically robust. Our process provides a method that reliably yields superconducting interconnects suitable for use with superconducting qubits.

\end{abstract}

\vspace{2pc}
% \noindent{\it Keywords}: qubit, superconducting, interconnect, flip chip, quantum integrated circuits

% \noindent{\it pacs}: 74.78.Fk superconducting heterostructures, 85.25.Hv superconducting integrated circuits,  85.25.Hv microelectronics superconducting circuits, 74.78.Fk superconducting multilayers,74.81.Fa Superconducting wire networks,84.71.Mn Superconducting wires, fibers, and tapes, 74.78.-w superconducting thin films
%
% Uncomment for keywords
%\vspace{2pc}
%\noindent{\it Keywords}: XXXXXX, YYYYYYYY, ZZZZZZZZZ
%
% Uncomment for Submitted to journal title message
%\submitto{\JPA}
%
% Uncomment if a separate title page is required
%\maketitle
% 
% For two-column output uncomment the next line and choose [10pt] rather than [12pt] in the \documentclass declaration
%\ioptwocol
%

\section{Introduction}

As superconducting qubit technology grows beyond one dimensional chains of nearest neighbor coupled qubits \cite{barends2014superconducting}, arbitrarily sized two dimensional arrays are a likely next step towards both surface code error correction and more complex high fidelity quantum circuits \cite{PhysRevA.86.032324}.  While prototypical two dimensional arrays have been demonstrated \cite{corcoles2015demonstration, riste2015detecting, PhysRevLett.117.210505}, the challenge of routing control wiring and readout circuitry has thus far prevented the development of high fidelity 3\,x\,3 or larger qubit arrays. For example, frequency tunable Xmon transmon qubits on the interior of a two dimensional array would require capacitive coupling to four nearest neighbor qubits and a readout resonator as well as individual addressability of an XY drive line and an inductively coupled flux line \cite{PhysRevLett.111.080502}. Routing these control wires with a single layer of base wiring and crossovers is not scalable beyond a few-deep array of qubits.  Multilayer fabrication with embedded routing layers is a natural solution \cite{PhysRevB.81.134510}, but integrated dielectric layers on a qubit wafer introduce additional decoherence to the qubits \cite{doi:10.1063/1.2898887}.  This individual addressability problem can be solved by separating the device into two chips, a dense wiring chip that allows for lossy dielectrics and a pristine qubit chip with only high coherence materials.  Combining these two chips to form a hybrid device provides the advantages of both technologies.

A hybrid device is composed of a ``base substrate'' bonded to a ``top chip.'' Hybridization allows for improved impedance matching between chips as compared to wirebonds and the close integration of incompatible fabrication processes. A qubit hybrid would also benefit from the availability of straightforward capacitive, inductive, or galvanic coupling of electrical signals between the base substrate and top chip through the use of parallel plate capacitors and coupled inductors.  Hybrid devices have become ubiquitous in the semiconductor industry, finding applications in everything from cell phones to the Large Hadron Collider \cite{BROENNIMANN2006303}. Cryogenic applications are fewer; bolometer arrays for submillimeter astronomy \cite{1278145, HILTON2006513} and single flux quantum devices \cite{0953-2048-25-10-105012, 4982605} have utilized this technology. Low resistance cryogenic bump bonds \cite{2017arXiv170604116R, 2017arXiv170502435M} and superconducting bump bonds that proximitize normal metals have also been fabricated \cite{2017arXiv170802219O}. Here we present a novel bump bond metal stack up consisting of all superconducting materials with the intent of achieving maximal flexibility in designing flux tunable qubit circuits where mA control currents are necessary.

In order to maintain compatibility with our existing qubit architecture, 
bump bond interconnects for a superconducting qubit hybrid must meet these requirements:
\begin{enumerate}
\item Bumps must be compatible with qubit fabrication (e.g., aluminum on silicon).
\item If interconnects will be used in routing control signals (rather than just as ground plane connections and chip spacers), fabrication yield must be high. e.g., With a 99.9\% yield, a device with 700 interconnects on control lines would yield all lines (0.999$^{700}$ =) 50\% of the time. 
\item Interconnects must continue to perform electrically and mechanically after cooling from 300\,K to 10\,mK.
\item  Bonding must be accomplished at atmospheric pressure without elevated process temperatures to avoid altering Josephson junction critical currents through annealing \cite{doi:10.1116/1.3673790}. 
\item Interconnects must superconduct to provide a lossless connection between chips and avoid local heating.
\item The critical current of the interconnects must exceed 5 mA to enable applications in current-biased flux lines.
\end{enumerate}

To satisfy condition (i) above and to extend our wire-routing capabilities through known multi-layer techniques, bumps must provide a connection between aluminum wiring on both the base substrate and top chip.  This design consideration will allow us to connect our qubit fabrication to a dense, multi-layer, wire routing device based on standardized complementary metal-oxide-semiconductor (CMOS) fabrication techniques.  Known bump bonding materials that also superconduct include indium and various soldering alloys. Indium is a natural choice because high purity sources are readily available, it can be deposited in many $\mu$m thick layers by thermal evaporation, it has a relatively high critical temperature of 3.4\,K, and room temperature indium bump bonding is an industrially proven technology \cite{datta2004microelectronic}.  However, since aluminum and indium form an intermetallic \cite{wade1973chemistry}, under bump metalization (UBM) it is necessary to act as a diffusion barrier.  Fortunately, titanium nitride, fulfills our UBM requirements as it is a well known diffusion barrier (used in CMOS fabrication) with a T$_{c}$ as high as 5.64\,K and has also been shown to be a viable high-coherence qubit material \cite{doi:10.1063/1.4813269, 0953-2048-27-1-015009}.

\section{Device fabrication and layout}

\begin{figure}[h]
\begin{center}
\includegraphics[width=0.8\textwidth]{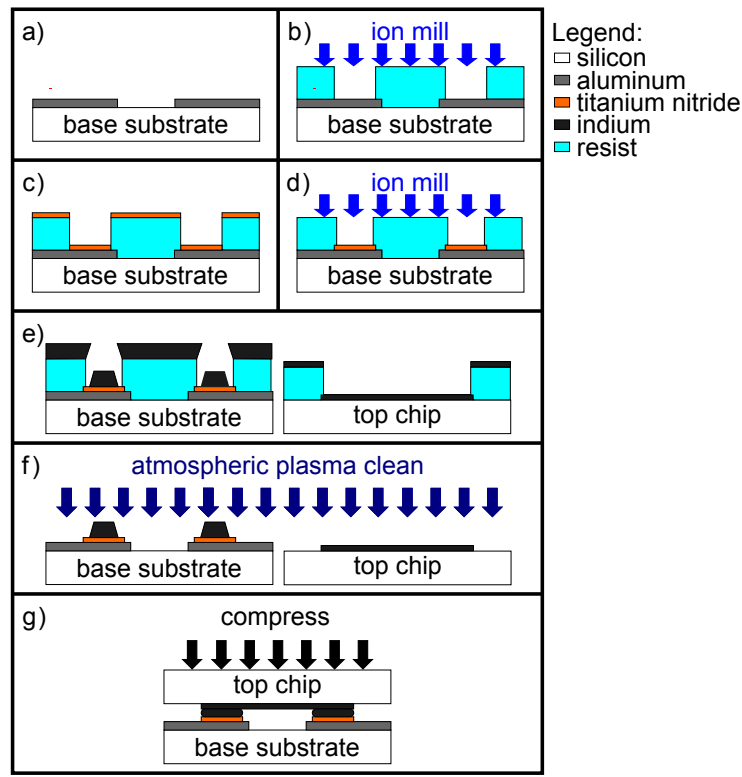}
\caption{Hybrid fabrication process;\,(a-d) describe steps specific to the base substrate and (e-g) are common to both the base substrate and top chip. a) On a silicon substrate, a base electrode is defined in 100 nm of e-beam evaporated aluminum by a BCl$_3$\,+\,Cl$_2$ plasma dry etch. b) The native aluminum oxide is removed by an ion mill at locations defined by lift-off resist. c) In the same vacuum chamber as b), 50-80 nm of titanium nitride is sputter deposited from a pure titanium source in argon and nitrogen partial pressures. d) After lift-off of the titanium nitride and patterning new resist, oxide and contaminants are removed from the titanium nitride by an ion mill at locations defined by lift-off resist. e) In the same vacuum chamber as d), 2-10\,$\mu$m of indium is deposited by thermal evaporation on both the base substrate and top chip. f) After lift-off of the indium, an atmospheric plasma is used to clean and passivate the surface of both devices a few minutes before bonding. g) The base substrate and top chip are aligned and compressed together at room temperature to complete the hybrid.\label{fig:fab_process}}
\end{center}
\end{figure}

Figure \ref{fig:fab_process} shows a minimal, qubit compatible, asymmetric bump bond process used here for DC characterization.  The base substrate has a full aluminum/titanium nitride/indium metal stack and, for simplicity, the top chip has just a single layer of indium wiring (which allowed us to avoid the complication of processing with two die sizes in every fabrication run while still testing all the necessary metal interfaces). In this case, as current flows between the base substrate and top chip, it passes through one aluminum/titanium nitride interface, one titanium nitride/indium interface, and one indium/indium interface. Actual qubit hybrids would be symmetric, with aluminum wiring and titanium nitride UBM on both chips, which adds one aluminum/titanium nitride interface and one titanium nitride/indium interface to the metal stack for each interconnect.

For the base substrate, we first blanket deposit 100\,nm of aluminum through e-beam evaporation--the same base wiring material used in qubit fabrication \cite{doi:10.1063/1.4993577}.  The base wiring, shown in figure (\ref{fig:fab_process}a), is defined with optical lithography and a BCl$_3$\,+\,Cl$_2$ plasma dry etch (although lift-off defined aluminum base wiring has been used with similar results).  Then, (\ref{fig:fab_process}b) titanium nitride pads are defined in lift-off resist and the device is placed into a sputter chamber where an \textit{in situ} ion mill (see \ref{appendix:ion_mill} for ion milling parameters) removes the native oxide from the aluminum (\ref{fig:fab_process}b) before titanium nitride is reactively sputtered in argon and nitrogen partial pressures (\ref{fig:fab_process}c).  After titanium nitride lift-off, the indium pillars are defined in lift-off resist and, then (\ref{fig:fab_process}d), in a third vacuum chamber, another in situ ion mill (\ref{appendix:ion_mill}) is used to remove oxide and contaminants from the titanium nitride surface, before depositing indium in a thermal evaporator with the substrate cooled to 0\,$^\circ$C (\ref{fig:fab_process}e).  Also shown in (\ref{fig:fab_process}e) is the single layer of indium lift off used to define indium wiring on the top chip--this may be done in the same or different indium deposition as the base substrate$\textsc{\char13}$s indium layer. For the devices we characterized here, we deposited 5\,$\mu$m of indium on the substrate and 2\,$\mu$m of indium on the top chip.

After both the base substrate and top chip have been fabricated, an atmospheric plasma surface treatment (with a mix of of hydrogen, helium and nitrogen) is used to remove surface oxide and passivate the surface of the indium a few minutes before the two chips are bonded together (\ref{fig:fab_process}f).  This surface treatment is critical to making good indium-to-indium contact during bonding without reflowing the indium \cite{6248801}.  We then flip over the top chip, align the two devices, and compress the dies together using a SET FC-150 flip-chip bonder (\ref{fig:fab_process}g). Bonding is performed at room temperature with a typical bonding force of 10-20\,N per mm$^2$ of bump area for 15\,$\mu$m diameter bumps (2-5\,grams/bump), which results in a compression of roughly 40-60\% the total height of the two indium depositions. Inspection with an edge gap tool indicates that typically the tilt between the base substrate and top chip is parallel within $\pm$\,0.5\,mRad, and inspection with an infrared microscope indicates that the xy alignemnt is typically within $\pm$\,2\,$\mu$m.

Choosing an appropriate bump geometry is subject to several constraints.  First, it is desirable to have a chip-to-chip separation of at least several microns so that the impedance of a 2\,$\mu$m wide, 50\,$\Omega$ coplanar waveguide transmission line is not dramatically changed by the presence of an overhead ground plane. Providing sufficient separation allows designs to be insensitive to the final chip-to-chip separation and for a smooth impedance transition as transmission lines travel under the edge of the top chip. In order to achieve a desired separation of 2-10\,$\mu$m post-compression, 2-10\,$\mu$m of indium must be deposited on both the base substrate and top chip. When depositing such thick layers of material, especially a high mobility material like indium, sidewall deposition can result in a considerable constriction of the bump feature size. 15\,$\mu$m diameter bumps were chosen as they have a width to height aspect ratio of 3:2 at the thickest intended bump height; for more information on thick indium deposition see \ref{appendix:fab}.  Secondly, the titanium nitride UBM footprint must be large enough so that, after compression, indium does not contact aluminum directly. Given the post-compression alignment accuracy of our flip chip bonder ($\pm$\,2\,$\mu$m) and an expected 50\% compression, we find that 30\,$\mu$m square titanium nitride pads are sufficient for 15\,$\mu$m diameter indium pillars. 

\begin{figure}[h]
\begin{center}
\includegraphics[width=1.0\textwidth]{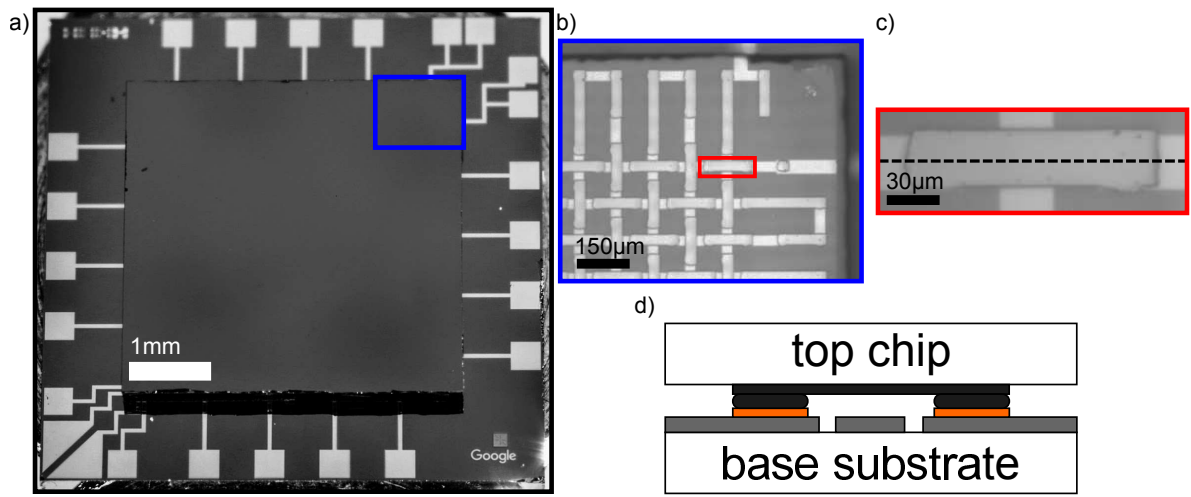}
\caption{Design of the bump bond DC characterization hybrid. a) Photograph of a hybrid device with a 6\,mm\,x\,6\,mm base substrate and a 4\,mm\,x\,4\,mm top chip. b) Infrared micrograph looking through the top chip of the hybrid device.  The woven pattern of test circuit can be seen, and bumps are located on either side of the crossings to connect the base wire from the base substrate to the top chip and back.  c) Zoomed in infrared micrograph of a single indium bar on the top chip with interconnects at either end.  d) Cross-sectional diagram of the device along the dotted line in c).\label{fig:device_cross_section}}

\end{center}
\end{figure}

The devices characterized here consist of a 6\,mm\,x\,6\,mm base substrate and a 4\,mm\,x\,4\,mm top chip shown in Figure \ref{fig:device_cross_section}.  In order to electrically characterize a large number of interconnects, we place 1620, 15\,$\mu$m diameter, circular indium bumps on the base substrate and 30\,$\mu$m\,x\,150\,$\mu$m indium bars on the top chip to connect pairs of bumps into a series chain of 1620 chip-to-chip interconnects.  At each end of the chain, and every 90 interconnects along the chain, we wire bond to pads on the perimeter of the chip.  This wiring configuration allows us to make four-wire resistance measurements by applying an excitation current to any 90 interconnect subsection (or number of subsections of the device) while measuring the voltage across that subsection/s with other leads.  Each section of 90 interconnects consists of three rows or columns that extend across the entire top chip, spread over an area of roughly 2\,mm$^{2}$.  By weaving these rows and columns of together, as shown in figure \ref{fig:device_cross_section}b, we are able to ascertain whether or not electrical failures are spatially correlated.  For instance, if one subsection arranged in the rows fails to superconduct or has a suppressed critical current, but none of the columns show the same behavior, it is likely that there are no spatially correlated failures.  However, if one section of rows and one section of columns fails, then the intersection indicates a region of interest for failure analysis such as electron energy loss spectroscopy (EELS), focused ion beam (FIB) cross sections, post-shear inspection, or inspection with an optical or infrared microscope.

\section{Electrical characterization}

We perform low temperature four-wire electrical measurements in an adiabatic demagnetization refrigerator (ADR) down to 50\,mK using a lock-in amplifier, ammeter, source measure unit (SMU) and a matrix switch to rapidly characterize a large number of devices. Twisted pair wiring and shielding is used to reduce parasitic coupling between the current excitation leads and voltage sense leads. Common mode voltage correction is implemented with the matrix switch which also allows us to quickly switch between measurements.  For a detailed look at the measurement system as well as the the resistance and critical current measurements discussed below, see \ref{appendix:bounding_resistance}.  

\begin{figure}[h]
\begin{center}
\includegraphics[width=1.0\textwidth]{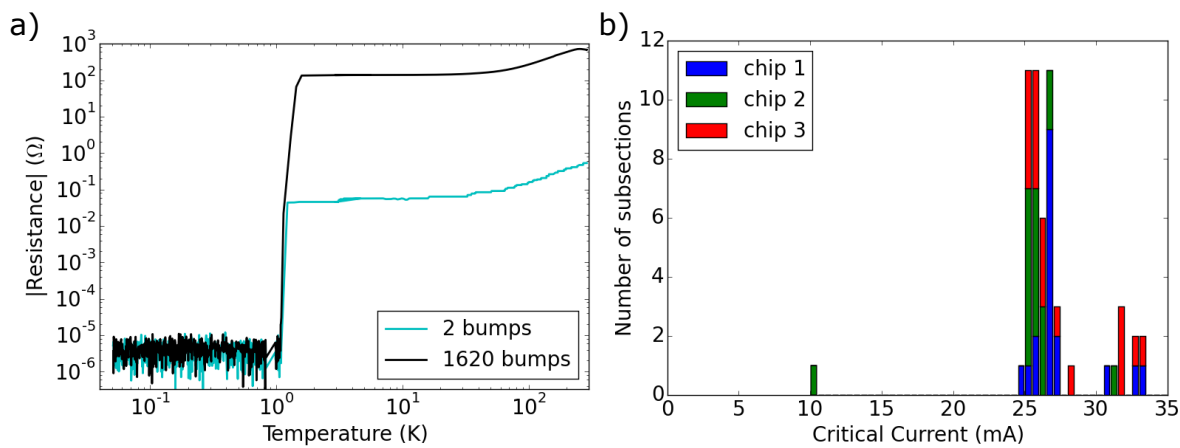}

\caption{Electrical device characterization. a) Typical four-wire resistance measurement versus temperature for a chain of 1620 interconnects and a 2 interconnect test structure on the same device from room temperature to 50 mK.  A superconducting transition can be seen at 1.1\,K where the resistance of both the 1620 and 2 interconnect structures fall to a few $\mu\Omega$.  For the 1620 long chain, this measurement demonstrates a superconducting resistance more than 7 orders of magnitude lower than its normal state resistance at 3\,K. b) Histogram of critical currents for each of the eighteen 90-interconnect subsections on three different chips. The average critical current is 26.8\,mA with $>$98\% of the subsections above 24.5\,mA \label{fig:results}}

\end{center}
\end{figure}

This setup allows us to make a resistance measurement of the device in its superconducting state.  Using common mode compensation and the lock-in amplifier with a several mA sinusoidal test current, we are typically able to bound the resistance of a series chain of 1620 interconnects to be less than 5\,$\mu\Omega$ below 1.1\,K, which is an average resistance of 3\,n$\Omega$ per interconnect. Figure \ref{fig:results}a shows a typical resistance versus temperature curve for a full 1620 interconnect chain and a 2 interconnect test structure on the same device.  At 1.1\,K we observe a clear transition to a superconducting state when the resistance of 1620 interconnects in series falls more than 7 orders of magnitude to a few $\mu\Omega$. The resistance measured below 1.1\,K is roughly the same for both 1620 interconnects and the 2 interconnect test structure which indicates that this measurement is likely limited by system parasitics or measurement electronics rather than by an actual resistance or the inductance of the device. In figure \ref{fig:results}b we use a SMU to assess the critical current of each of the eighteen 90 interconnect subsections on three hybrid devices. The average critical current for each subsection is 26.8\,mA, with a number of subsections above 30\,mA and a single subsection with a suppressed critical current of 10.3\,mA. This data represents 4860 interconnects, 100\% of which superconduct with a critical current above 10\,mA. Furthermore, at least 98\% of the interconnects have a critical current above 24.5\,mA. Since there was only one section of rows (and no columns) with a suppressed critical current, it is likely that a single interconnect could be responsible for the lower critical current. The high yield of this process and lack of spatially correlated failures indicate that parallel interconnects can be used to further increase the critical current and/or to serve as precautionary redundant connections (though we yielded 100\% on these 3 test devices and have had similar yields across several generations of test devices). The average room temperature resistance of these 90 interconnect subsections is 47.7\,$\Omega$ with a standard deviation of 2\,$\Omega$ indicating reasonable bump uniformity.  Typically we find that a room temperature resistance $<$1\,$\Omega$/interconnect (including the aluminum and indium base wiring used to chain them together) indicates that the flip chip bonding was successful. We find that insufficient compression or a bad material interface results in a resistance higher than 1\,$\Omega$ per interconnect.

\section{Mechanical characterization}

Several mechanical tests were performed on a different generation of hybrids consisting of a 10\,mm\,x\,10\,mm substrate and a 6\,mm\,x\,6\,mm square chip.  These devices had about four thousand 20\,$\mu$m diameter circular bump bonds spread fairly evenly over the 36\,mm$^2$ area of the top chip. In order to characterize the mechanical strength of these interconnects, we performed destructive die shear strength tests (in accordance with MIL-STD-883) in which a force is applied to the edge of the top chip, parallel to the face of the chip (e.g., force was applied in the plane of the page as the chip is shown in figure 2a), until the top chip separates from the substrate.  Four devices were tested; three separated at 35\,N and one exceeded the limits of the tool at 49.9\,N, all of which are more than sufficient to ensure that devices are robust enough for handling. Finally, thermal cycling was performed on a device that had been previously confirmed to be fully superconducting below 1.1\,K.  One hundred thermal cycles from -80\,$^{\circ}$C to 45\,$^{\circ}$C were performed with a 23 minute dwell at both -80\,$^{\circ}$C and 45\,$^{\circ}$C and a 20\,$^{\circ}$C/min ramp rate for transitions.  After 100 thermal cycles (and unknown conditions during round-trip ground shipping to our off-site lab) the sample was cooled back down to 50\,mK.  All interconnects on the device still remained superconducting, although the critical current was reduced to 1-5\,mA in most subsections down from 20-25\,mA in the initial characterization of this device. The reason for the reduced critical current is not known, but it is worth noting that, in a more typical use case, the devices measured in figure \ref{fig:results} were cycled from room temperature to 50\,mK and back as many as three times in our ADR (approximately 0.2$^{\circ}$C/min average warming/cooling rate) with no measurable impact on the critical current.

\section{Conclusion}
The flip chip hybrid devices we have developed offer a viable solution to control signal routing in two-dimensional high-coherence circuits. These interconnects, consisting of a titanium nitride diffusion barrier and indium bumps, serve as electrical interconnects between two planar devices with aluminum wiring.  This fabrication process opens the door to the possibility of the close integration of two superconducting circuits with each other or, as would be desirable in the case of superconducting qubits, the close integration of one high-coherence qubit device with a dense, multi-layer, signal-routing device.  Furthermore, these interconnects have a typical critical current above 25\,mA which is an order of magnitude larger than the largest typical DC control currents used to flux-tune superconducting qubits.  Limited by the aluminum, these bumps are fully superconducting below 1.1\,K, and below this critical temperature, we are able to estimate the resistance of each bump to $<$\,3\,n$\Omega$.  These high yield, mechanically robust, and high critical current electrical interconnects are ready to be implemented into more complex circuits including two dimensional arrays of nearest neighbor coupled flux-tunable superconducting qubits.

\subsection*{Acknowledgments}
This work was supported by Google. C. Q. and Z.C. acknowledge support from the National Science Foundation Graduate Research Fellowship under Grant No. DGE-1144085. Devices were made at the UC Santa Barbara Nanofabrication Facility, a part of the NSF funded National Nanotechnology Infrastructure Network. Materials characterization performed by the Google Failure Analysis lab. The authors would also like to thank Eric Schulte for sharing his wealth of experience.

\section*{References}
\bibliographystyle{unsrt}
\bibliography{refs}

\newpage
\appendix
\section{Electrical characterization measurement setup}
\label{appendix:bounding_resistance}

\begin{figure}[h]
\begin{center}
\includegraphics[width=1.0\textwidth]{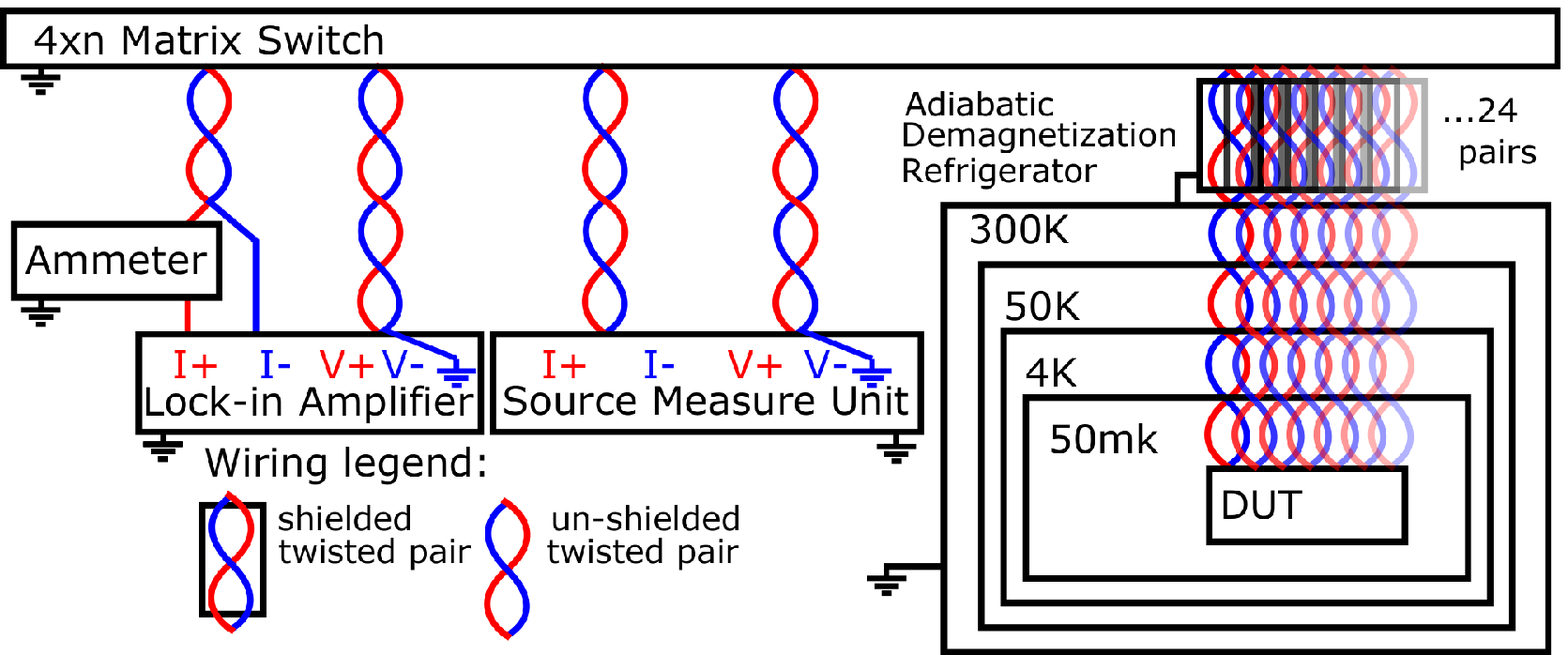}

\caption{Schematic of the measurement setup. A 4 by n matrix switch is used to route two sense and two excitation lines from various measurement equipment, including a lock-in amplifier and a source measure unit, to the bump bond devices.  Both the measurement equipment and DUT are connected to columns of the matrix switch and the rows are used to connect any column to any other column. The measurement equipment chassis are all grounded to a common surge protector. The twisted pair shielding is grounded at the ADR, and floating at the matrix switch. The four-wire measurement ground is provided by the negative excitation terminal of either lock-in amplifier or source measure unit. \label{fig:wiring}}

\end{center}
\end{figure}

Bounding the resistance of a suspected superconducting device requires the ability to very accurately measure the excitation current and the resulting voltage drop across the device.  Even when care is taken to use appropriate signal wiring and grounding, as shown and described in Figure \ref{fig:wiring}, DC based measurements are subject to thermoelectric voltages, broadband noise, and measurement ranges optimized for non-zero resistance materials where finite voltages are expected. An AC excitation and a digital lock-in amplifier can be used to mitigate these effects, but the resulting measurement is not without difficulties. A lock-in amplifier implements mixing and filtering to extract the signal amplitudes both in phase and 90-degrees out of phase with a reference tone at the specific reference frequency. Both the in phase voltage (V$_x$) and quadrature voltage (V$_y$) across a device may be extracted if the sinusoidal excitation is used as the lock-in amplifier reference.

\begin{figure}[h]
\begin{center}
\includegraphics[width=1.0\textwidth]{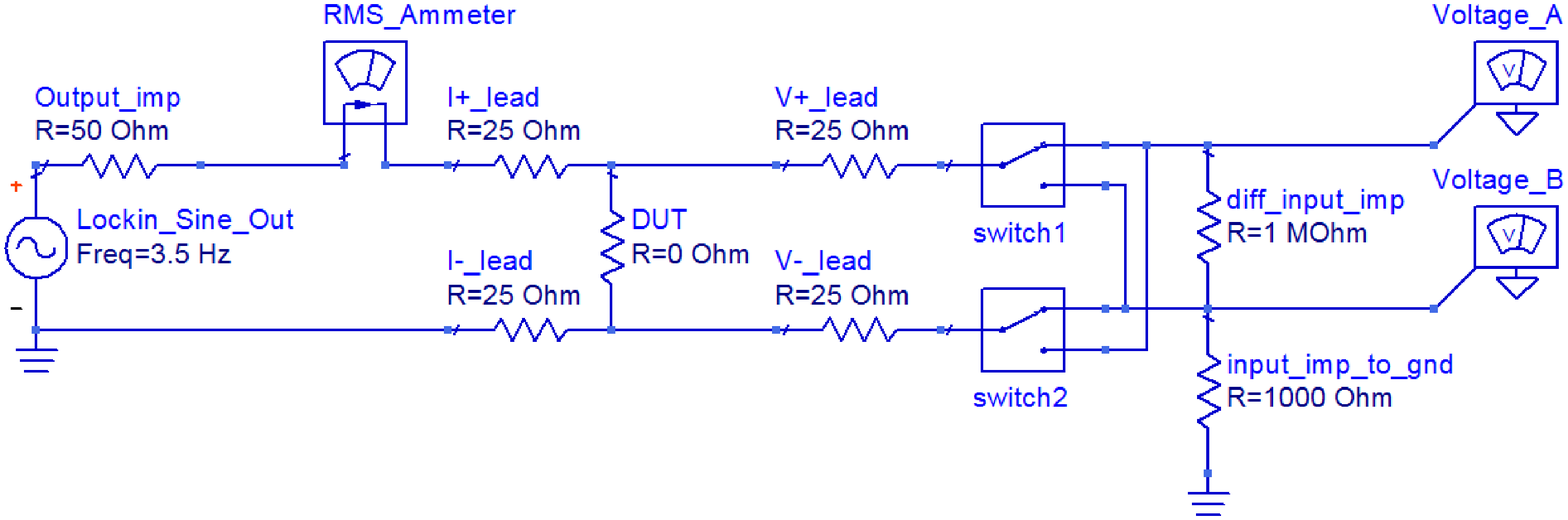}
\caption{A model of the 4-wire measurement circuit used to perform the bounding resistance measurement.\label{fig:four_wire}}
\end{center}
\end{figure}

Figure \ref{fig:four_wire} shows a model of the 4-wire measurement circuit used to perform a bounding resistance measurement.  The voltage excitation signal is an adjustable frequency sine-wave generator with a 50\,$\Omega$ output impedance provided by the lock-in amplifier, a Stanford Research Systems model SR830.  An ACrms ammeter, a Keysight Technologies model 34461A, is placed in-line with the positive voltage lead of the sine-wave generator to measure the excitation current, which is set by the amplitude of voltage waveform and the approximately 50\,$\Omega$ lead resistance in the I$_+$ and I$_-$ leads.  This lead resistance is dominated by the niobium titanium wiring used in our cryostat all the way from 300\,K to the 50\,mK stage and varies by 10-20\% channel-to-channel.  This lead resistance variation is why the excitation current is measured directly with the ammeter rather than inferring it from the excitation voltage.  

It is important to note that this wiring configuration results in a common-mode voltage at the sample approximately equal to half of the excitation voltage due to the voltage divider created by the excitation leads.  The lock-in amplifier used here has a common mode rejection ratio (CMRR) of 100 dB meaning that common mode (CM) voltages may leak into the differential voltage measurement attenuated by 10$^5$.  Without further common mode compensation (and independent of the excitation voltage) the 50\,$\Omega$ lead resistance and CMRR specification would limit the measurement accuracy as follows.  The excitation current is approximately equal to the excitation voltage divided by the total of the voltage source output impedance and the sum of the two excitation leads:

\begin{equation}\label{eq1}
{\rm I_{ex} = V_{ex}/R_{lead+source}}
\end{equation}

Since the excitation leads are approximately equal, the common mode voltage on both sense leads will be approximately V$_{ex}$/2 and the lock-in amplifier’s CMRR specifies how much of this voltage may leak into its differential voltage measurement:

\begin{equation}\label{eq2}
{\rm V_{cm leakage} = V_{cm} * CMRR = \frac{V_{ex}}{2} * CMRR = V_{ex} * 5 * 10^{-4}}
\end{equation}

Combining~\ref{eq1} and~\ref{eq2} we find that the common mode leakage and lead resistance would limit our measurement to a minimum of 500\,$\mu\Omega$.

\begin{equation}\label{eq3}
{\rm R_{min} = \frac{V_{cm leakage}}{I_{ex}} = \frac{V_{ex}}{2} * CMRR *(\frac{V_{ex}}{R_{leads+source}})^{-1} = 500\,\mu\Omega}
\end{equation}

\begin{figure}[h]
\begin{center}
\includegraphics[width=1.0\textwidth]{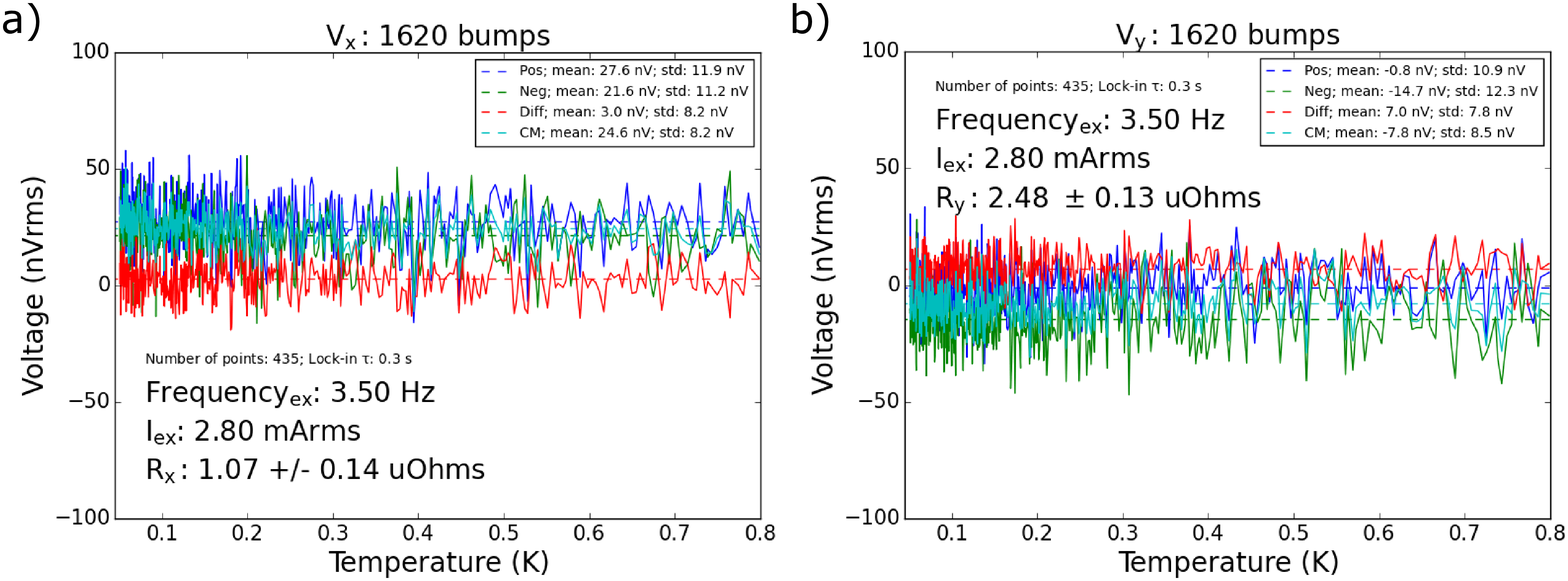}
\caption{Typical Vx and Vy voltages traces with positive and negative lead polarity.  The differential voltage computed from the difference of the positive and negative polarity measurements is near zero and the common mode voltage computed from the sum is consistent with the lock-in amplifiers CMRR specification of 100dB (we usually see 110-140 dB). \label{fig:cmr}}
\end{center}
\end{figure}

Compensation for this common mode voltage leakage is accomplished by taking two voltages measurement using the switches shown in figure \ref{fig:four_wire}.  While holding the excitation signal constant, the switches are used to reverse the polarity of the voltage sense leads and two measurements are recorded:

\begin{enumerate}
\item Differential Voltage\_A-Voltage\_B + common mode leakage, and
\item -(Differential Voltage\_A-Voltage\_B) + common mode leakage
\end{enumerate}
The sum of these two measurements is two times the common mode leakage and the difference is two times the differential voltage of interest.  Figure \ref{fig:cmr} shows these two voltage measurements as well as the computed common mode (CM) and differential voltages for the V$_x$ and V$_y$ signals measured across a 1620 bump structure.  This data confirms that the lock-in amplifier is meeting both its common mode rejection specification of $>$100\,dB, as well as its input noise specification of 6\,nV/$\rm\sqrt{Hz}$--since a 0.3\,s time-constant was used, the input noise should be $<$\,11\,nV$_{rms}$.

Since the lock-in amplifier is making an AC measurement, care must be taken to make sure that parasitic inductances and capacitances do not affect the measurement. Firstly, it is very important to use twisted pair wiring for at least one pair of, and preferably both, the sense +/- leads and the excitation +/- leads (grounded shielding should also be used where possible to further reduce mutual inductances of the sense and excitation leads and to further reduce electromagnetic noise pick up). By utilizing twisted pairs in this configuration, the mutual inductance between these leads is reduced considerably--without twisted pairs it is easy to end up with a several $\mu$H or more mutual inductance between the sense and excitation leads which will end up looking like an in-phase voltage (or real resistance). Secondly, the voltage signal from the inductance of the sample as well as other parasitic inductors and capacitors should be proportional to the frequency of the excitation voltage. To reduce the impact of such signals and parasitics, measurements were made using a low frequency excitation signal, typically $<$\, 10\,Hz.  Finally,  we typically find that the resulting differential voltage is proportional to both the excitation voltage as well as the excitation frequency indicating that the signal we are measuring is due to a system parasitic and is not just electronics noise.  The frequency dependence in particular hints that this load is primarily not resistive, but even in taking the conservative approach of assuming it is all resistive, this measurement system is able to limit the resistance of a series chain of 1620 bumps to be several $\mu\Omega$.  The fact that the differential voltage measured across 1620 bumps is the same as for 2 bumps is further evidence that the measurement is dominated by a cabling parasitic and not an actual resistance in the sample, or even the sample’s inductance.

Measuring the critical current of these interconnects in an Adiabatic Demagnetization Refrigerator (ADR) required some optimization to run efficiently.  An ADR uses a helium compressor to cool a superconducting magnet and sample stage down to 3-4\,K.  Then, to cool the sample down to 50\,mK, the current in the superconducting magnet is ramped up over 10-15 minutes to align magnetic dipoles in a salt crystal. After a 30-45 minute soak time in this magnetic field, the salt crystal and sample stage are thermally disconnected from the rest of the system and the magnetic field from the superconducting magnet is ramped down over 10-15 minutes.  When the magnetic field approaches zero, the dipoles in the salt crystal begin to mis-align, pulling heat out of the system and cooling the sample stage down to about 50 mK. If too much heat is added to the system then the 60-75 minutes magnet cycle must be repeated to cool back down. The superconducting devices tested here have a critical current $>$\,25\,mA, and once a subsection is driven normal by exceeding the critical current, the sample stage of the cryostat heats up from 50\,mK to 3\,K in a about a second if the current is not reduced.  In order to efficiently characterize many devices, care was taken to avoid unnecessarily heating the cryostat.

In order to limit the heat dissipation of the sample in the cryostat, a Keysight Technologies model B2901A source measure unit (SMU) was used. A SMU is a combination source (with a configurable current or voltage set point) integrated with a meter (configurable for current and/or voltage).  In this case, a current set point is used and the voltage across the sample is measured--as the current is increased, if the voltage across the sample jumps above the noise level, then the sample has transitioned to a normal state. SMUs are fast, accurate, and offer a number of features that enabled us to make hundreds of critical current measurements in a single ADR magnet cycle.  Firstly, this SMU offers pulsed operation where the source provides a timed current pulse and the measurement aperture is synchronized to occur just after the excitation has settled.  We found that we were able to achieve good results using just a 2\,ms long current pulse with a 0.4\,ms measurement aperture window.  Furthermore, this SMU offers a voltage protection feature where the source terminals are physically disconnected inside the unit with a relay if the source compliance condition (compliance voltage in this case) is reached.  Since a superconducting material is being measured, the compliance voltage at the device should be 0\,V.  A compliance limit of 5\,mV was set, and we found that these results were in good agreement with measurement where we did not pulse the excitation.

\newpage
\section{Material and interface characterization}
\label{appendix:fab}

Aluminum is deposited using e-beam evaporation in a vacuum chamber with a base pressure of 1e-7\,mBar. 100\,nm of aluminum is deposited at a rate of 1\,nm/s. Structures were patterned and etched using standard lithographic techniques and BCl$_3$\,+\,Cl$_2$ chemistry in an inductively coupled plasma etcher. (Other samples have yielded using both wet etches and lift-off defined structures.)

The titanium nitride under bump metalization (UBM) is used as a diffusion barrier between indium and aluminum as both are known to be very reactive metals \cite{wade1973chemistry}. To achieve a dense film with low oxidation and Tc above 3\,K, we employ a substrate bias during deposition \cite{6942188}. A 50-80\,nm titanium nitride film is grown using a reactive sputter (150\,W power) from a pure titanium target in 3\,mTorr of argon and nitrogen (48\,sccm and\,1.75 sccm flows, respectively). 

\begin{figure}[h]
\begin{center}
\includegraphics[width=0.75\textwidth]{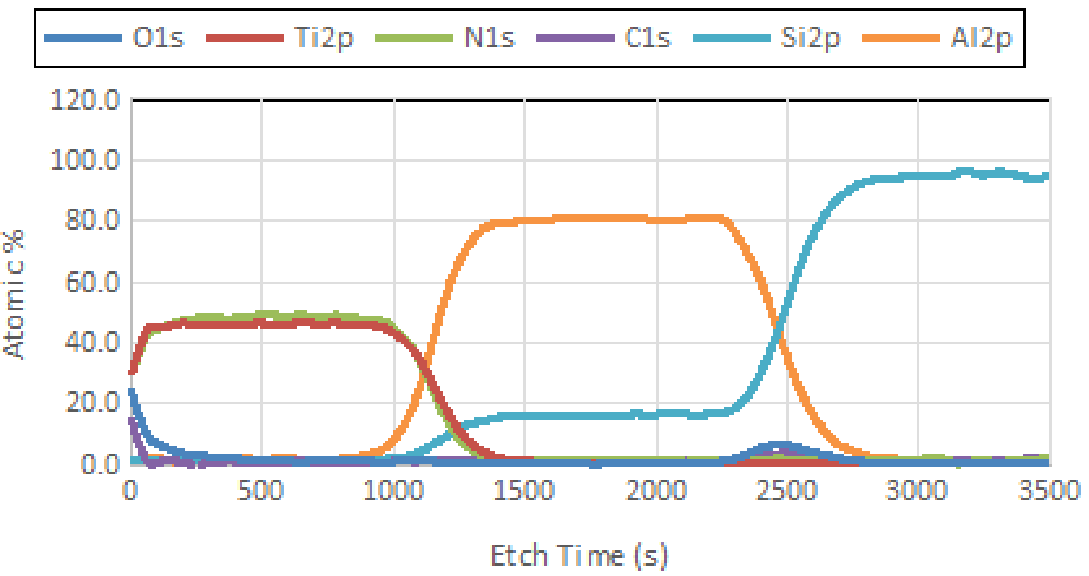}
\caption{XPS data for a layer of titanium nitride on aluminum on a silicon substrate.\label{fig:xps}}
\end{center}
\end{figure}

The resulting films are found to be nearly stoichiometric, but slightly nitrogen rich using X-ray photoelectron spectroscopy (XPS) (Figure \ref{fig:xps}) and Rutherford backscattering spectroscopy (RBS) (Table \ref{rbs}). Moreover interdiffusion of aluminum into the titanium nitride is absent.

\begin{table}[h]
\caption{\label{rbs}RBS data for a 320 {\AA} titanium nitride layer on 1000 {\AA} of aluminum on a silicon substrate.}
\footnotesize
\begin{tabular}{|c|c|c|c|c|c|c|c|c|}
\br
&\multirow{2}{*}{"RBS" Thickness [{\AA}]}&\multicolumn{6}{c|}{Atomic Concentration [at\%]}&\multirow{2}{*}{Assumed Density [at/cc]}\\
&&{\bf N}&{\bf Si}&{\bf Ti}&{\bf Al}&{\bf W}&{\bf Ar}&\\
\br
Layer 1&320&53.5&-&46.5&-&-&-&1.07E23\\
\mr
Layer 2&10&-&-&-&29.7&3.7&66.6&3.79E22\\
\mr
Layer 3&1000&-&-&-&100&-&-&6.02E22\\
\mr
Bulk &-&-&100&-&-&-&-&.5.00E22\\

\br
\end{tabular}\\
\end{table}

Since titanium nitride is employed as a diffusion barrier, it is deposited as square pads beneath the indium bumps. These pads are defined in lift-off, using a single layer of positive photoresist and a MIB (AZ Developer) developer to prevent etching and roughening of the underlying aluminum during developer rinse. To make good electrical contact between the titanium nitride and the underlying aluminum, the patterned aluminum wafer is ion milled in-situ before sputter deposition. Mill parameters are shown in  \ref{appendix:ion_mill} (120\,s mill time). 

\begin{figure}[h]
\begin{center}
\includegraphics[width=0.75\textwidth]{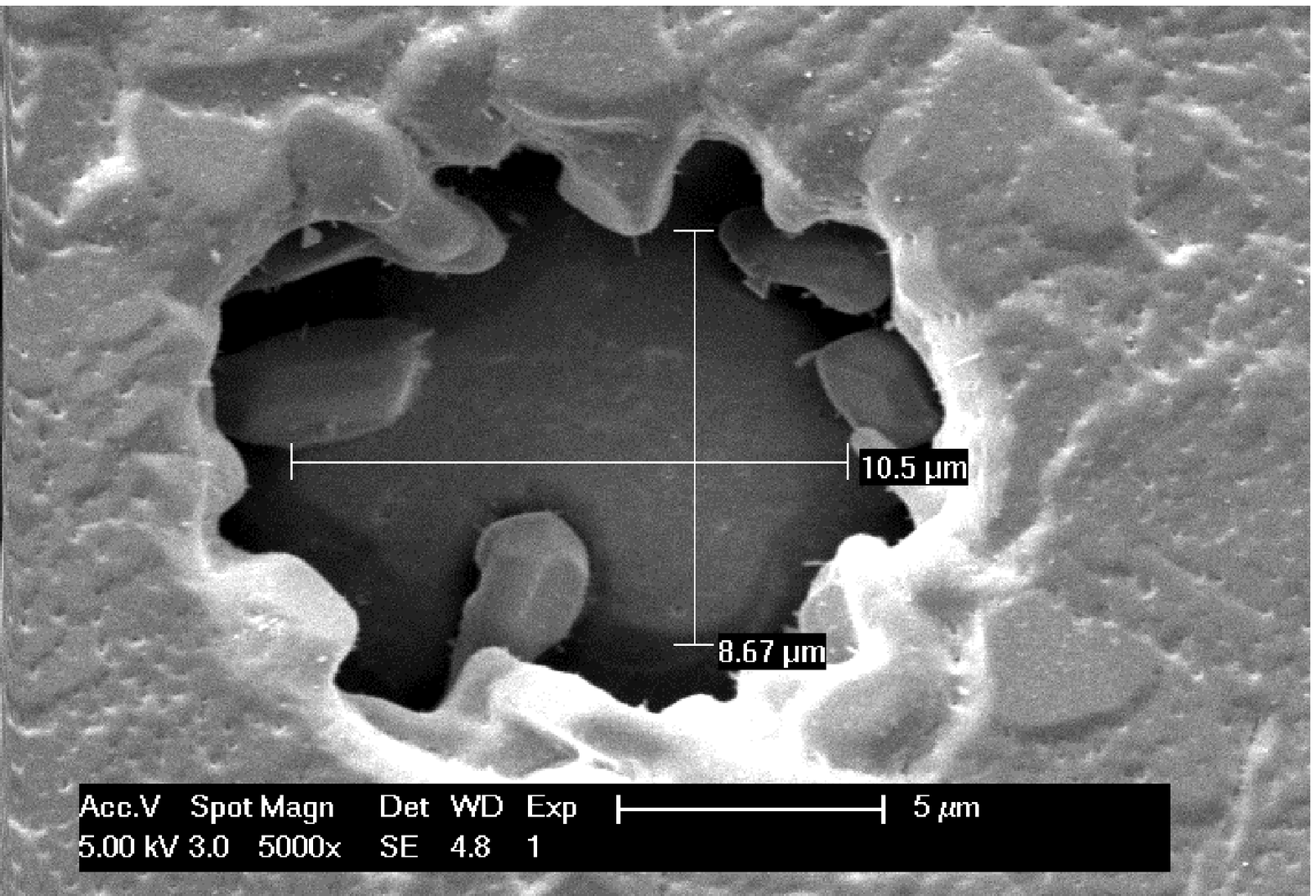}
\caption{SEM image of indium crystallite growth over a 15\,$\mu$m diameter hole during a slow ($<$1\,nm/sec) indium deposition.\label{fig:sem}}
\end{center}
\end{figure}

After the titanium nitride is lifted off, the wafer is patterned again using a lift-off polarity and a thick positive resist. Circular apertures are opened using a MIF developer (since the aluminum is encapsulated by corrosion resistant titanium nitride). The wafer is loaded into a thermal evaporator with a base pressure below 1E-7\,Torr. To remove any contaminants and insulating oxides, the wafer is ion milled \textit{in situ}, then allowed to cool on the water cooled chuck (held at 0\,C). Indium is deposited at rates exceeding 2 nm/s to prevent a constriction of the lithographically defined apertures by crystallite growth. Figure \ref{fig:sem} is a SEM of typical crystallite growth that occurs when slow deposition rates are used. Indium lift-off is performed in a heated NMP bath.   

\begin{figure}[h]
\begin{center}
\includegraphics[width=0.75\textwidth]{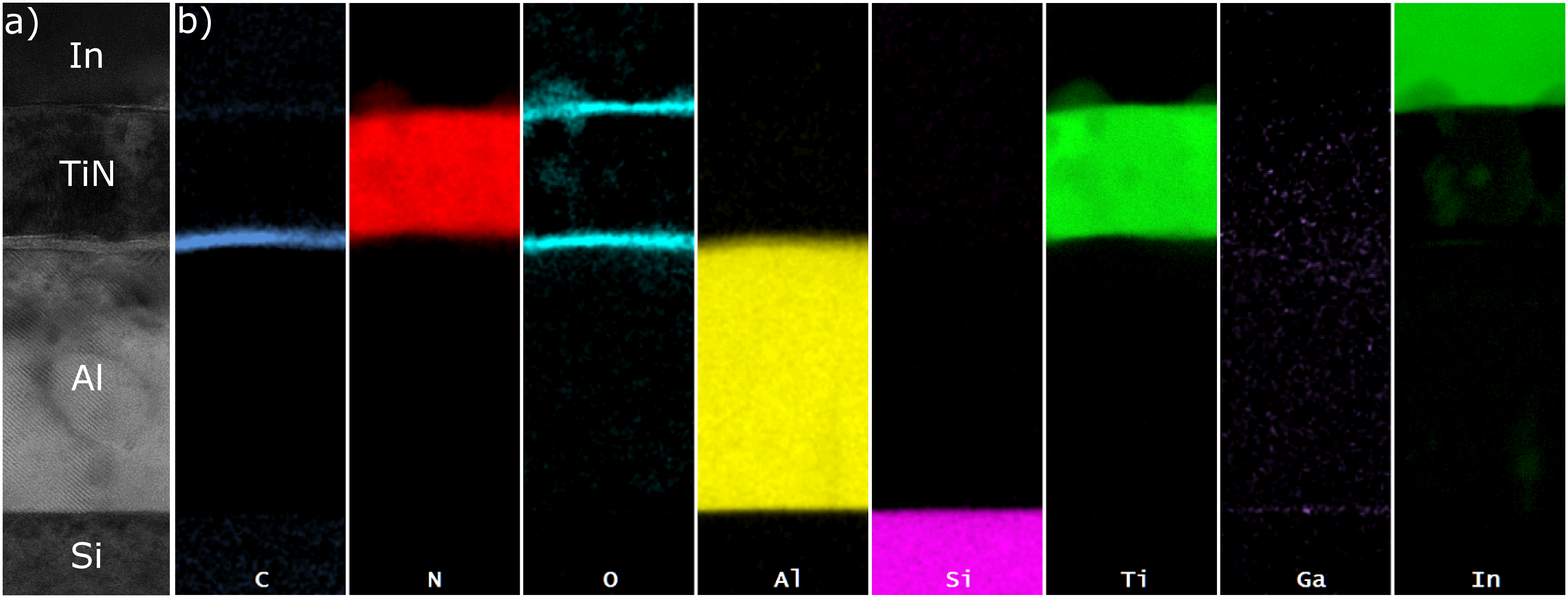}
\caption{Electron energy loss spectroscopy of a focused ion beam cross-section of one interconnect. a) SEM image of the focused ion beam cross section of an indium bump on a titanium nitride diffusion barrier with aluminum base wiring on a silicon substrate. b)  Electron energy loss spectroscopy of the sample shows in a). This confirms the titanium nitride to be a sufficient diffusion barrier as there is no indium to aluminum contamination.  Oxide contamination can be seen at both titanium nitride interfaces, but this does not seem to affect the critical current of these interconnects.  The carbon present at the aluminum/titanium nitride interface is due to redeposition of lift off photoresist during the ion mill of the aluminum a and the gallium present is from the focused ion beam used to cut the cross section.  \label{fig:eels}}
\end{center}
\end{figure}

The entire material stack up has been characterized using focused ion beam (FIB) cross sections and electron energy loss spectroscopy (EELS), as shows in figure \ref{fig:eels}, to determine the composition of the layers and, most importantly, their interfaces. Crucially, no indium-aluminum interdiffusion exists across the titanium nitride barrier. However, intermittent oxide contamination (up to 15\% by atomic percent) at the titanium nitride/indium interface and titanium nitride/aluminum interface has been measured on various samples, although this seems to have little affect on yield or critical current. 

\newpage
\section{Ion mill parameters}
\label{appendix:ion_mill}
\begin{table}[h]
\caption{\label{alion}\textit{In situ} ion mill parameters used to clean aluminum surface before depositing titanium nitride. Ion mill time is 120\,s.}
\footnotesize
\begin{tabular}{@{}lllllll}
\br
&Cathode&Discharge&Beam&Accelerator&Neutralizer&Emission\\
\mr
Voltage (V)&7.3&40.0&399&79&18.9&n/a\\
Current (A)&10.8&0.48&0.055&0.0031&17.2&0.118\\
\br
\end{tabular}\\
\end{table}
\normalsize

\begin{table}[h]
\caption{\label{inion}\textit{In situ} ion mill parameters used to clean titanium nitride surface before depositing indium. Ion mill time is 90\,s.}
\footnotesize
\begin{tabular}{@{}lllllll}
\br
&Cathode&Discharge&Beam&Accelerator&Neutralizer&Emission\\
\mr
Voltage (V)&9.3&40.0&600&120&10.4&n/a\\
Current (A)&14.9&1.52&0.12&0.004&0.0122&0.116\\
\br
\end{tabular}\\
\end{table}
\normalsize

\end{document}